\documentclass[pre,twocolumn,amsmath,amssymb,notitlepage,nofootinbib,aps,longbibliography]{revtex4-1}
\usepackage{dcolumn}
\usepackage{lipsum}
\usepackage{graphicx}
\usepackage{hyperref}
\usepackage{float}
\usepackage{bm}
\usepackage[T1]{fontenc}
\usepackage{color}
\usepackage{url}
\usepackage[bf]{subfigure}
\usepackage{rotating}

\begin{document}

\title{Patterns of authors contribution in scientific manuscripts}

\author{Edilson A. Corr\^ea Jr.$^1$}

\author{Filipi N. Silva$^2$}

\author{Luciano da F. Costa$^2$}

\author{Diego R. Amancio$^1$}
\email{diego@icmc.usp.br}

\affiliation{
    $^1$Institute of Mathematics and Computer Science, University of S\~ao Paulo, PO Box 668, 13560-970, S\~ao Carlos, Brazil\\
    $^2$S\~ao Carlos Institute of Physics, University of S\~ao Paulo, PO Box 369, 13560-970, S\~ao Carlos, SP\\}


\begin{abstract}
Science is becoming increasingly more interdisciplinary, giving rise to more diversity in the areas of expertise within research labs and groups. This also have brought changes to the role researchers in scientific works. As a consequence, multi-authored scientific papers have now became a norm for high quality research. Unfortunately, such a phenomenon induces bias to existing metrics employed to evaluate the productivity and success of researchers. While some metrics were adapted to account for the rank of authors in a paper (their position along the sequence of authors), many journals are now requiring a description of the specific roles of each author in a publication. Surprisingly, the investigation of the relationship between the rank of authors and their credited contributions has been limited to a few studies. By analyzing such kind of data, here we show, quantitatively, that the regularity in the authorship contributions decreases with the number of authors in a paper. Furthermore, we found that the rank of authors and their roles in papers follows three general patterns according to the nature of their contributions, such as writing, data analysis, and the conduction of experiments. This was accomplished by collecting and analyzing the data retrieved from \emph{PLoS ONE} and by devising an entropy-based measurement to quantify the effective number of authors in a paper according to their contributions. The analysis of such patterns confirms that some aspects of the author ranking are in accordance with the expected convention, such as the fact that the first and last authors are more likely to contribute more in a scientific work. Conversely, such analysis also revealed that authors in the intermediary positions of the rank contribute more in certain specific roles, such as the task of collecting data. This indicates that the an unbiased evaluation of researchers must take into account the distinct types of scientific contributions.
\end{abstract}

\maketitle

\section{Introduction}
\label{sec:introduction}
Scientific publications have been consolidated as the most efficient means of communication and dissemination of discoveries/findings in science, as well as being an indicator of success/productivity of a given researcher, evidenced by quantitative measures, including the $h$-index~\cite{hirsch2005index,Schreiber2015150,Amancio2012427} and its variants (such as $m$-index~\cite{hirsch2005index} and $g$-index~\cite{Egghe2006}). Even though these measures are used to evaluate researchers, their indiscriminate use is still controversial given the existence of problems such as lack of informative context~\cite{wendl2007h} and the presence of possible bias towards nationality, gender and age~\cite{mishra2008citations,kelly2007h}. In addition, the use of citations based indexes  are oftentimes very dependent on authors' research field, which makes comparisons of individuals prominence across distinct disciplines a hard task~\cite{Waltman2013833,anauati2016quantifying,RuizCastillo2015102,Hutchins029629,Waltman2011,Viana2013371}.
Another limitation in evaluating authors according to the number of citations motivated by their papers is that the contributions of authors in each work is mostly overlooked, as in traditional indexes such as the h-index all authors usually receive the same credit~\cite{sekercioglu2008quantifying,zhang2009proposal}. In recent years, however, some attempts have been proposed to take into account additional information such as the number of authors in papers~\cite{bornmann2008there} and the consideration of authors rank~\cite{tscharntke2007author}.

The evaluation of authors contribution in scientific articles has drawn attention in recent years, mainly because of its implication in the author credit system~\cite{greene2007demise} for quantifying authors roles in a scenario of increasing number of multi-authored papers~\cite{regalado1995multiauthor}. The growing interest in the quantification of authors role in every step of the research led several journals to adopt policies encouraging authors to include information regarding individual contributions. This is the case of journals such as \emph{Nature}\footnote{http://www.nature.com/nature}, \emph{Science}\footnote{http://www.sciencemag.org}, \emph{PNAS}\footnote{http://www.pnas.org}, \emph{PLoS One}\footnote{http://journals.plos.org/plosone/} and \emph{Journal of Informetrics}\footnote{http://www.journals.elsevier.com/journal-of-informetrics}.
This information allows for the quantification of unprecedented patterns, such as the variability of contributions, the relationship between rank (position in the list of authors) and the amount of contributions. More important than that, the knowledge of authors contribution may lead to a more informed quantification of researchers abilities, thus improving the author credit system as a whole.

In this paper we analyze the information describing the authors contributions present in research papers. Upon devising a simple framework to handle the contributions of authors as a bipartite network, we propose a measurement to quantify the \emph{effective number} of authors in a paper. We found that the effective number of authors, in general, is a linear function of the total number of authors. Most importantly, we could identify that the deviation between the actual and the effective number of authors increases with the total number of authors. Using a measure to quantify the symmetry/diversity of contributions in papers, we also found out that the regularity of contributions decreases as the number of authors increases from 1 to 10. We also analyzed the nature of authors contributions and their relationships with the rank of authors. This study revealed the existence of three general patterns that governs how authors are ranked according to their specific roles in scientific publications.

This paper is organized as follows. Section \ref{methodology} describes the considered dataset and the framework to represent and evaluate the contributions of authors in papers. Section \ref{results} details the main results of this paper. Finally, in Section \ref{conclusion} we present the conclusions of the study.

\section{Methodology} \label{methodology}

In this section, we briefly describe the dataset used in our analysis and the pre-processing steps applied to obtain authors contributions in scientific articles. We also describe a measure to quantify the irregularity of individual contributions in papers in terms of the following quantities: the \emph{effective number} of authors; and the \emph{symmetry} of contributions.

\subsection{Database} \label{dataset}

In order to probe the authors contribution patterns in scientific papers, we created a dataset comprising articles retrieved from the \emph{PLoS One} journal. We particularly chose this journal because it spams many disciplines and provides parseable information concerning authors contribution in each paper. We also chose this journal because it provides a large number of papers, which makes the analysis less prone to undesirable small sample effects. We have retrieved about 80,000 articles published between 2006 and 2014. For each article, we recovered the authors' names and the full list of contributions. To investigate the relationship between contributions and rank, we also recovered the position of authors in the list of authors.

Concerning the retrieved information, only the list of contributions had to be processed. In \emph{PLoS ONE}, contributions are associated with the acronym of authors' names.  Because there is no clear rule about generating acronyms from names, authors may generate acronyms in distinct ways. To obtain an association between acronyms and actual names, the following procedure was adopted: (i)  we created an acronym for each author based on the capital letters of his/her name; (ii) we associated perfect matches between provided acronyms and the ones generated in step (i); (iii) if a perfect match is not obtained in the previous step, a similarity measure was used in order to associate acronyms. More specifically, we used the Tanimoto's distance because of its widespread use in comparing strings and acronyms~\cite{tanimoto58}.

To simplify our analysis, we first captured all possible contributions informed by authors. Then, these items were reduced to a smaller set comprising only the most frequent contributions. This set includes the following contributions, as informed by authors: (i) \emph{analyzed the data}, (ii) \emph{collected the data}, (iii) \emph{conceived the experiments}, (iv) \emph{performed the experiments}, (v) \emph{wrote the paper}, and (vi) \emph{revised the manuscript}. We have also clustered very similar contributions to one of the above mentioned items. The groups of equivalent or similar contributions are:
\begin{enumerate}
	
\item {\bf Analyzed the data}: this contribution also includes the following items: ``analyzed the data'', ``interpretated the data'', ``statistical analysis'', ``performed a statistical analysis'', ``interpreted the results'', ``data interpretation'' and ``contributed to the discussion''.
	
\item {\bf Conceived the experiments}: this contribution also includes the following items: ``conceived and designed the experiments'', ``designed the software used in the analysis'', ``designed the study'', ``designed the experiments'', ``conceived and designed the study'',

\item {\bf Performed the experiments}: this contribution refers to a single item: ``performed the experiments''.

\item {\bf Wrote the paper}: this contribution also includes the following items: ``wrote the paper'', ``wrote the manuscript'', and ``contributed writing the manuscript''.

\item {\bf Collected the data}: this contribution also includes the following items: ``contributed with reagent materials and analysis tools'' and ``collected the data''.

\item {\bf Revised the manuscript}: this contribution also includes the following items: ``revised the manuscript'', ``edited the manuscript'', ``reviewed the manuscript'', ``read and approved the final manuscript'', ``critical revision of the manuscript'', ``critically revised the manuscript'',  ``critical review of the manuscript'', ``revised the paper'', ``edited the paper'', ``reviewed and edited the manuscript'' and ``critically reviewed the manuscript''.

\end{enumerate}

Another pre-processing step applied to obtain relevant information from the original dataset concerns the consideration of institutions (universities, departments, etc) as collaborators.  Because  no specific contribution is provided for institutions, we have ignored all papers with this type of information. In addition, we have also disregarded all articles with no explicit information concerning authors contribution. In total 15,787 articles were removed from the analysis. The pre-processed dataset is available for download in our website\footnote{\emph{PLoS ONE} dataset:\\ \url{http://cyvision.ifsc.usp.br/patternsauthors}\\ (18 sep 2016)}.

\subsection{Effective number of authors and symmetry of contributions} \label{ena}

\begin{figure*}[]
\centering\includegraphics[width=0.7\linewidth]{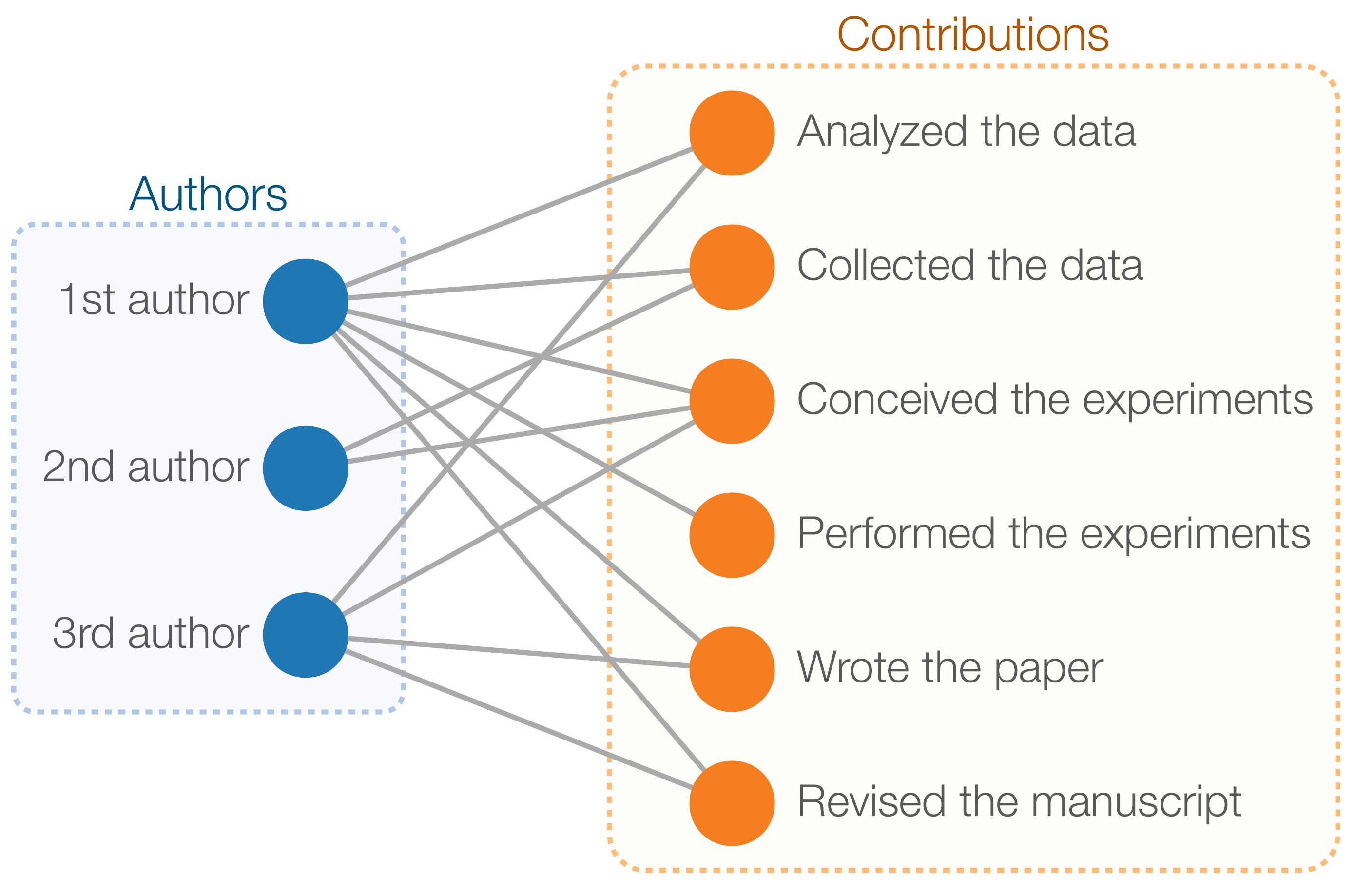}
\caption{\label{fig_network}Example of bipartite network representing the relationship between authors and contributions. The list of all possible contributions was obtained from the information provided by authors who published in the \emph{PLoS ONE} journal.  Note that the total amount of authors and particular contributions vary from article to article.}
\end{figure*}

The analysis of the contribution of a particular author can be accomplished in several ways~\cite{citestr}, as our dataset directly indicates which contributions an author may have, as shown in Figure~\ref{fig_network}. The information provided in each article of our dataset can be regarded as a bipartite network~\cite{Newman:2010:NI:1809753}, i.e. a network with links established only among nodes belonging to distinct groups. As illustrated in Figure \ref{fig_network}, the bipartite network derived from each paper establishes links between authors and its possible contributions.

\begin{figure*}[]
\centering\includegraphics[width=0.65\linewidth]{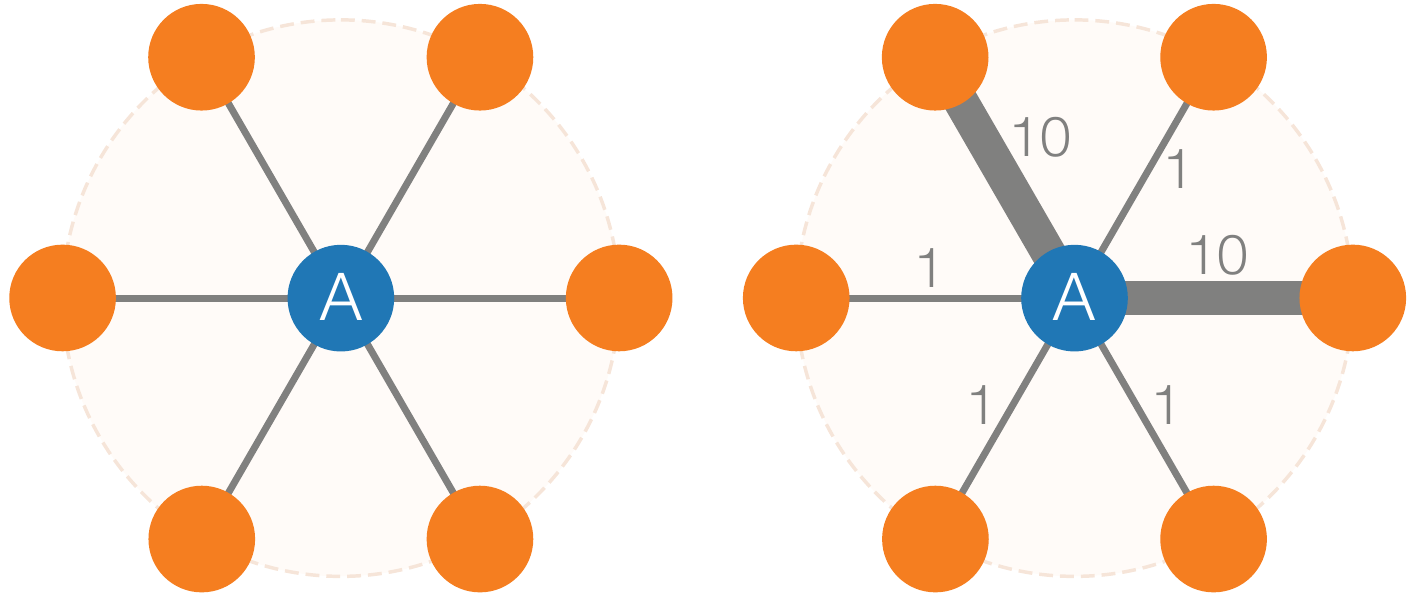}
\caption{\label{exacc}Quantification of the accessibility  (a) in a network whose links possess the same weight; and  (b) in a heterogeneous network. In (a), a random walker leaving node A access all neighbors
with the same probability. As such, the accessibility is the same as the node degree, i.e. $\alpha_A = k_A = 6$. In (b), two nodes receives most of the random walkers leaving node A. Therefore, the effective number of neighbors drops to $\alpha_A=3.52$, according to equation \ref{eq.acc}. Similar preferential random walks used applied to bibliographic analysis have been proposed e.g. in~\cite{6137404,Amancio2012}.}
\end{figure*}

A interesting feature of bipartite networks representing contributions is the regularity of contributions in a paper.
For simplicity's sake, we consider that all contributions are equally important and, for this reason, we study regularity in terms of the diversity of contributions made by authors. To measure how regular is the contribution of authors in papers, we used a measure inspired in the \emph{accessibility} concept, a centrality measurement employed in network science~\cite{Travencolo200889,1742-5468-2015-3-P03005}. This measurement was originally devised~\cite{Travencolo200889} to capture the \emph{effective number} of accessed nodes when an agent performs  a random walk from a starting node.
Differently from traditional measurements, such as the node degree ($k_i$),  the accessibility use network features that go beyond the simple static network topology. The accessibility uses the probability of reaching each adjacent node to quantify the effective number of neighbors. Let $p_{ij}$ the probability of a random walker to go from node $i$ to node $j$. The accessibility $\alpha_i$ is given by
\begin{equation} \label{eq.acc}
	\alpha = \exp \Big{(} - \sum_j p_{ij} \log p_{ij} \Big{)},
\end{equation}
where the sum is performed for all neighbors such that $p_{ij}>0$. In other words, the accessibility is the exponential of the entropy of the distribution of $p$. As a consequence, the maximum value that the accessibility can obtain is $\alpha_i^{(\textrm{max})}=k_i$, which occurs when $p_{i1}=p_{i2}=\ldots=p_{ik_i}$~\cite{Travencolo200889}. Whenever the distribution of $p$ is uneven, the random walkers tend to access preferentially specific nodes, thus decreasing \emph{effective number} of accessed nodes~\cite{Travencolo200889}. This effect is illustrated in Figure~\ref{exacc}. In (a), the central node has an accessibility value $\alpha_A = k_A = 6$, as all six neighbors are equally accessed in a random walk (edges weight are equally distributed). Conversely, in (b), the distribution of visits is more concentrated in two neighbors, because the strength of their links are much higher than the strength linking A and other neighbors. Owing to the large deviation in the distribution of  $p$ in this scenario, the effective number of neighbors drops to $\alpha_A = 3.52$, as defined in equation \ref{eq.acc}.

The proposed measure for quantifying the effective number of authors in a scientific article takes its maximum value when all authors contributes equally to the study, a scenario similar to the network configuration in Figure \ref{exacc}(a). In a paper, if a given author performs most of the contributions and many authors only perform a single task, the effective number of authors drops to a value close to 1. Consequently,  the effective number the authors takes low values whenever the distribution of authors contributions is uneven.
The contribution of authors is quantified using the following representation of a bipartite network. Let $\mathbf{B}$ be the matrix storing the relationship between authors and contributions,  i.e.
\begin{equation} \label{eq:clas}
 B_{ij} = \\
 \begin{cases}
	 	 1, & \textrm{if the $i$-th author performs the $j$-th contribution;}   \\
 	          0, & \textrm{otherwise.}
 \end{cases}
\end{equation}
In the example provided in Figure \ref{fig_network}, $B_{2j}=1$ only for $j=\textrm{``collected the data''}$ and $j=\textrm{``conceived the experiments''}$.
The total contribution of a given author $a$ is given by
\begin{equation} \label{eqca}
	c_a = \frac{\sum_j w_j B_{aj}}{\sum_i \sum_j w_j B_{ij}},
\end{equation}
where $w_j$ is the weight associated to the $j$-th contribution. In this study, we consider all contributions equally important and, for this reason, we set $w_j = 1$ for all contributions.
Note that the contribution of each author ($c_a$) defined in equation \ref{eqca} ranges in the interval $0 \leq c_a \leq 1$. Therefore, we can measure the diversity of the distribution of $c_A$ using the entropy $\mathcal{H}$ for all authors authors in the set of authors $\mathcal{A}$:
\begin{equation}
	\mathcal{H} = - \sum_{a\,\in\,A} c_a \log c_a.
\end{equation}
As proposed in the definition of the accessibility in equation \ref{eq.acc}, the definition of the effective number of authors according to the diversity of contributions can then be defined as
\begin{equation}
	\mathcal{N} = e^\mathcal{H} = \exp \Bigg{(} - \sum_{a \in \mathcal{A}} c_a \log c_a \Bigg{)}.
\end{equation}

In our analysis, we also probed how contributions varies across authors using a normalization of the \emph{accessibility} defined in equation \ref{eq.acc}. The normalized \emph{accessibility}, referred to as \emph{symmetry} of contributions, takes a range of values restricted in the interval $[0,1]$ and is defined as

\begin{equation} \label{eq.sym}
	\sigma = \frac{e^\mathcal{H}}{n_A} = \frac{1}{n_A}   \exp \Bigg{(} - \sum_{a \in \mathcal{A}} c_a \log c_a \Bigg{)},
\end{equation}
where $n_A$ is the total number of authors in the paper. Note that $\sigma$ is a symmetry measure because it reaches its maximum value when all authors contribute equally to the paper.

\section{Results and discussion}\label{results}

\begin{figure}[h]
\centering\includegraphics[width=0.95\linewidth]{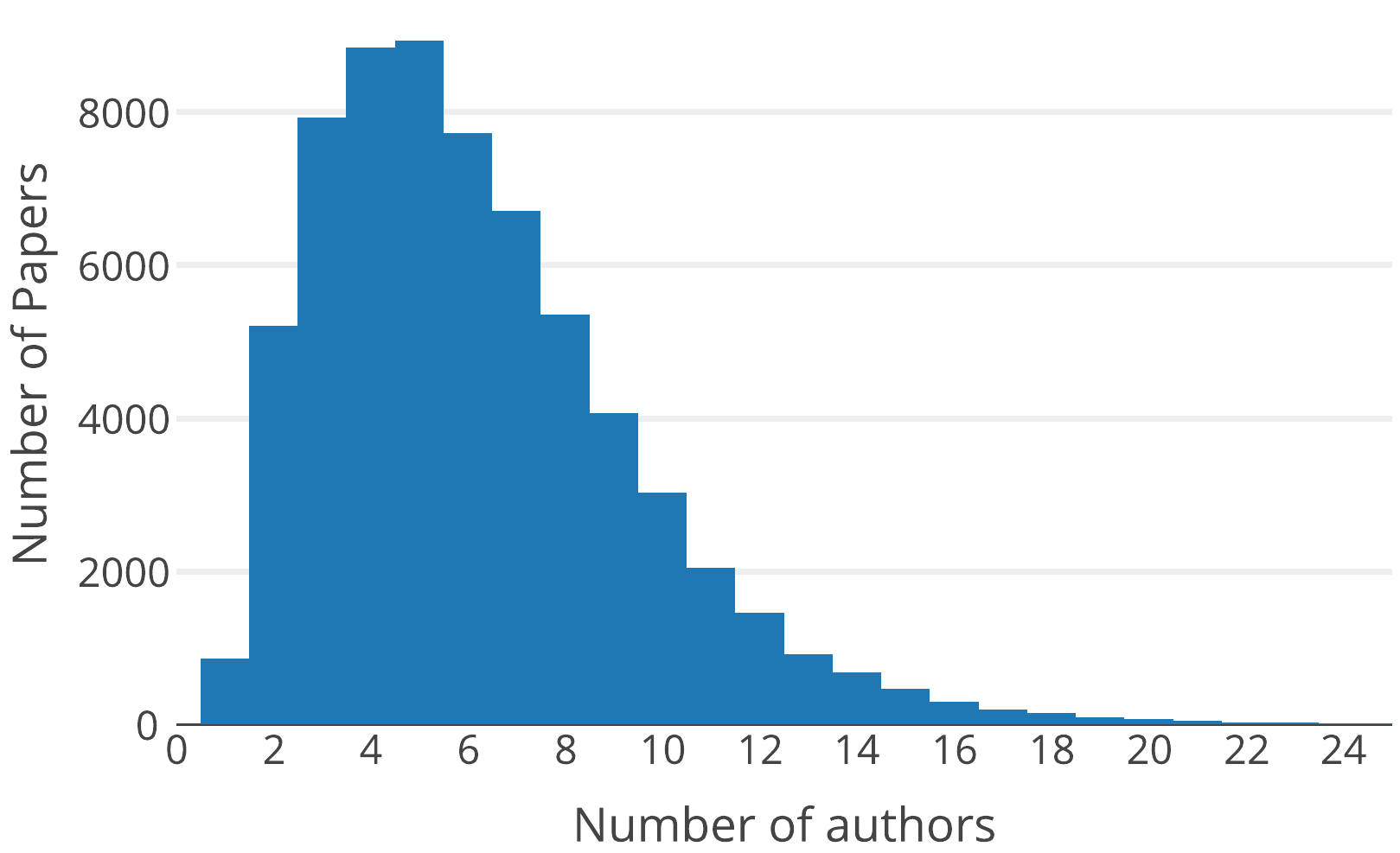}
\caption{\label{natu}Histogram of number of authors in the \emph{PLoS ONE} dataset reveals a irregular distribution: most of the papers includes the collaboration of 2-9 authors. Interestingly, we also observe a significant number of papers including more than $10$ authors, which is in accordance with recent reported trends~\cite{10.15200/winn.141832.26907}. }
\end{figure}

We start our analysis by quantifying specific statistics of our dataset (see Section \ref{dataset}). The distribution of the number of authors per paper is shown in Figure \ref{natu}.
Note that only a few papers contain a single author, while the most common scenario are papers from 2 to 10 authors. This is perhaps the common scenario in most academic fields, where shared authorship contributes to a wider view on the academic research. However, some criticism has been put on paper authored by many authors, as this pattern of collaboration may not reflect major author contributions. In fact, multi-authored papers may contribute to boost authors individual performance, as authors with minor contributions in several papers may easily benefit from the attributed authorship~\cite{sekercioglu2008quantifying}. Such bias towards authors with minor contributions may even occur when other factors such as number of citations is analyzed. To clarify specific contributions in multi-authored papers, individual contributions have been described in several journals. In our dataset, the main contributions (according to their frequency in articles) are shown in Table \ref{my-label}. Almost half of all authors contributed to analyze the data. A similar percentage was obtained for the contributions ``perform the experiments'' and ``conceived the experiments''. A slight smaller percentage of authors contributed to write the manuscript. Finally,  about one third of all authors collected the data for the study. Because the contribution ``revised the manuscript'' is not significant, we have disregarded this information in our experiments.

\begin{table}[h]
\centering
\caption{Frequency of appearance for each type of contribution considering all paper of the dataset. Each author was counted as a distinct occurrence, even if they appear in more than one paper of the dataset. }
\label{my-label}
\begin{tabular}{|l|c|}
\hline
{\bf Contribution} & {\bf Fraction} \\
\hline
Analyzed the data & 49.97\% \\
Performed the experiments &  48.47\%\\
Conceived  experiments & 45.33\%\\
Wrote the paper & 42.08\% \\
Collected the data & 34.07\% \\
Revised the manuscript & 0.30\%\\
\hline
\end{tabular}
\end{table}

\begin{figure}[h]
\centering\includegraphics[width=0.80\linewidth]{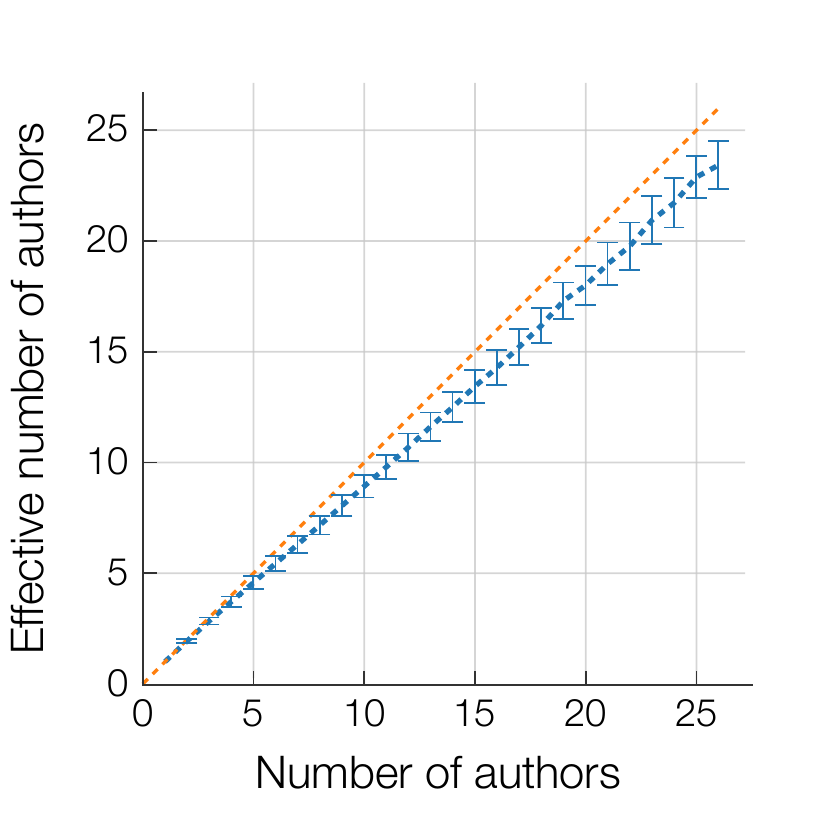}
\caption{\label{reta}Effective number of authors ($\mathcal{N}$) as a function of the actual number of authors ($n_A$). Because in some cases some authors contribute more than others, the effective number of authors is lower than the total amount of authors in the paper. The highest deviations occur for the papers authored by many authors.}
\end{figure}

To investigate how individuals contributions varies in papers, we used the \emph{effective} number of authors ($\mathcal{N}$) as a measure of variability, as defined in Section \ref{ena}. Since the value of $\alpha$ varies according to the total (actual) number of authors ($n_A$), we show separately the values of $\mathcal{N}$ for each $n_A$. In Figure \ref{reta}, the red dotted line is the reference curve $\mathcal{N} = n_A$ and the blue circles denote the points observed in our dataset. When $n_A=1$, $\mathcal{N}=1$, as one should expect from equation \ref{eq.acc}. When $n_A$ increases, the \emph{effective} number of authors also increases, thus confirming a strong correlation between these quantities. The largest deviations between these quantities (i.e. $n_A - \mathcal{N}$) were found for the papers authored by many authors.  Note e.g. that, in general, contributions in papers authored by 22 authors are so irregular that one can consider that, in average, contributions are \emph{effectively} performed by 20 authors. Considering this set of articles comprising 22 authors, the paper with the most irregular distribution of contributions has an effective number of authors of
of only about $17$ authors. Despite these discrepancies, we can conclude that in a typical paper authored by 1-10 authors, the difference between $n_A$ and $\mathcal{N}$ is very small, as the differences in amount contributions performed in these cases is not significant.

\begin{figure}[!htbp]
\centering\includegraphics[width=0.80\linewidth]{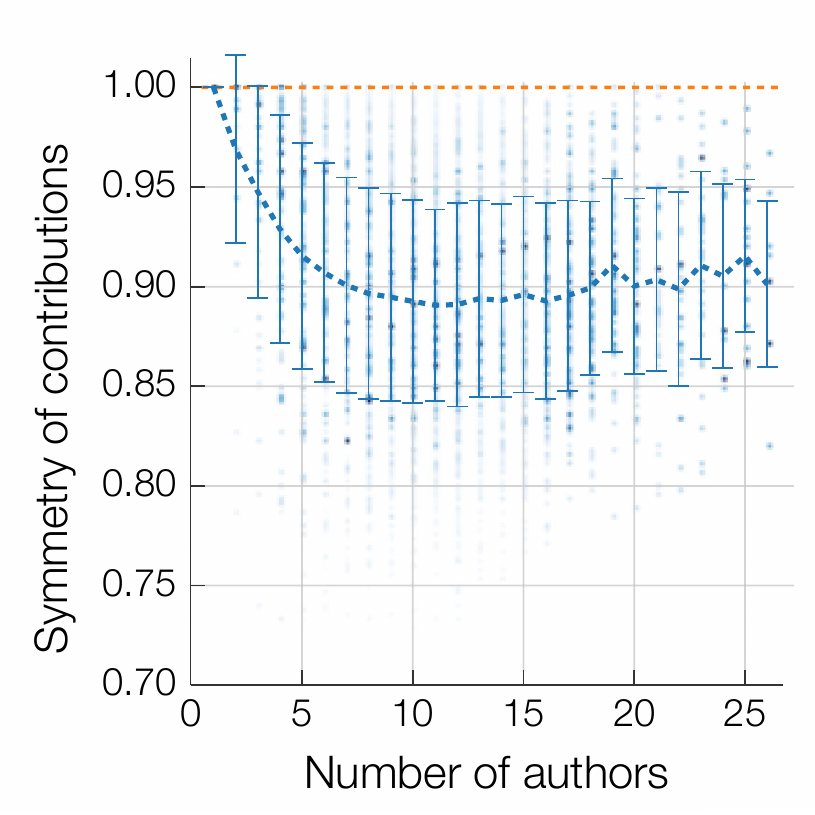}
\caption{\label{figsym}Symmetry of contributions in papers ($\sigma$) as a function of the total number of authors. For papers comprising 1 to 10 authors, the symmetry is inversely proportional to the number of authors. For papers comprising more than 10 authors, the symmetry of contributions is practically constant.}
\end{figure}

The irregularity of contributions was also investigated in terms of the \emph{symmetry} of contributions, as defined in equation \ref{eq.sym}. In Figure \ref{figsym}, we show, for each value of $n_A$, the corresponding value of symmetry.  The red dotted line represents the curve obtained by linking the points representing the average symmetry obtained for each $n_A$. As expected, the symmetry takes its maximum value when $n_A=1$. The average symmetry  monotonically decreases when the number of authors goes from $n_A = 1$ to $n_A = 10$. This means that contributions become more irregular with the number of authors. However, when more authors run in the authorship list, the symmetry of contributions remains almost constant. Owing to the large number of papers authored by 2-10 authors, outliers are more common in this subset of articles. This is evident if we note that values of symmetry $\sigma \leq 0.80$ are not frequent in articles with more than 10-15 co-authors.

A recurrent problem in assigning credit to researchers concerns the choice of adequate rankings according to authors specific contributions.
Although using the rank of authors may lead to a good approximation of the real credit that a author deserves, there is no straightforward manner to perform ranking~\cite{secondthoughts}. Some studies suggest good practices to rank authors, such as giving important credits to first and last authors, with distinct roles~\cite{tscharntke2007author}. It has been suggested that first authors are the ones making the greatest contributions, while last authors are the ones who designed and proposed the study~\cite{tscharntke2007author}. While this assertive remains true across most of disciplines, there is no major consensus on how to rank intermediary co-authors. For this reason, in this study, we also analyzed the contributions as a function of ranks to identify if there is a implicit factor leading the organization of rankings according to contributions. We also investigate how first and last authors compare to intermediary authors in terms of specific contributions.

\begin{figure*}[!htbp]
 \centering
   \includegraphics[width=0.95\linewidth]{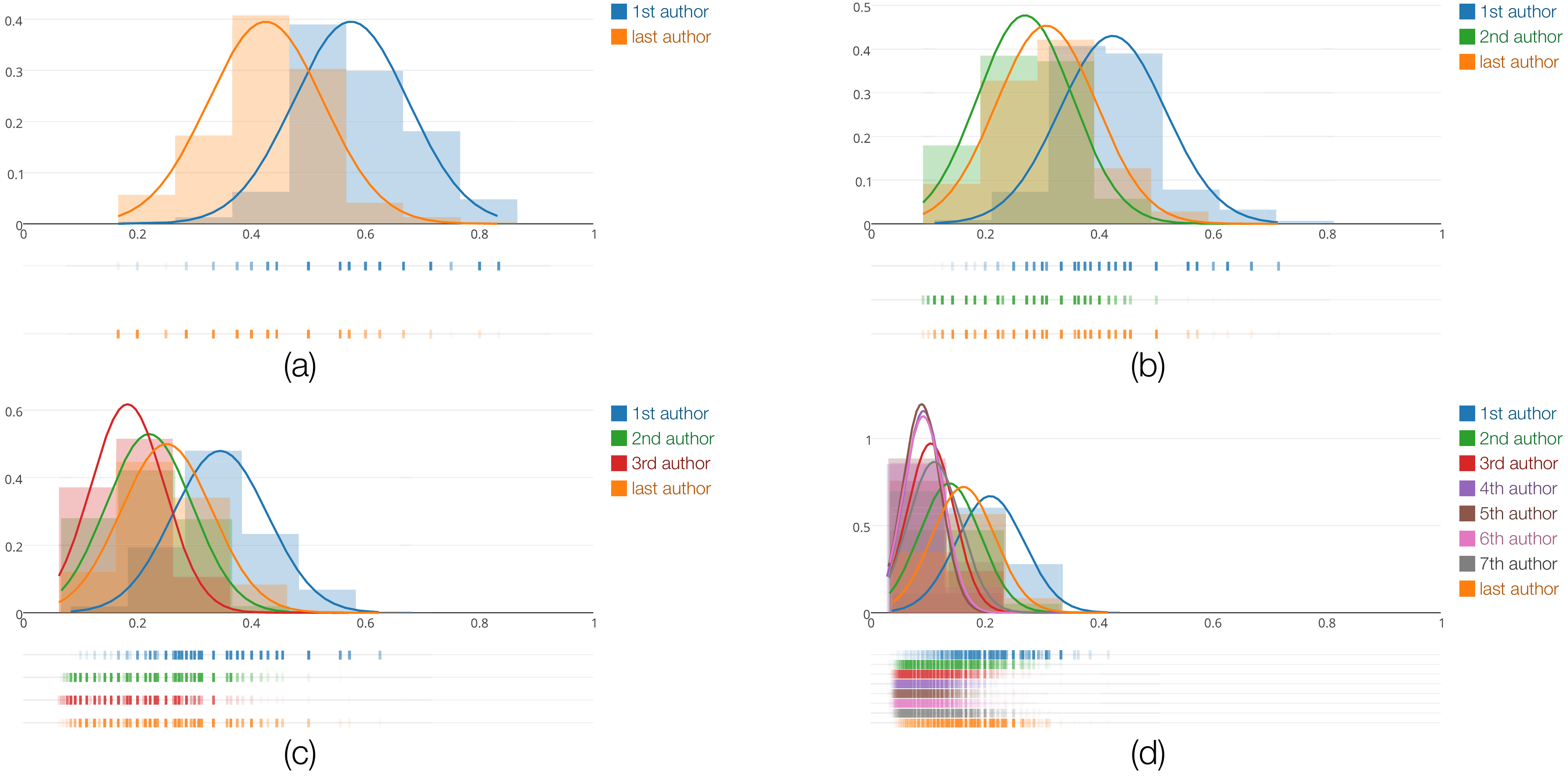}
\caption{\label{contribution-multi}Contributions of authors in scientific papers as a function of their rank. The results are shown considering the following number of authors: (a) 2; (b) 3; (c) 4; and (d) 8. Additional figures are shown in the Supplementary Information (Figure S1). Interestingly, both first and last authors are the ones with the largest number of distinct contributions.}
\end{figure*}

\begin{figure*}[!htbp]
\centering\includegraphics[width=0.75\linewidth]{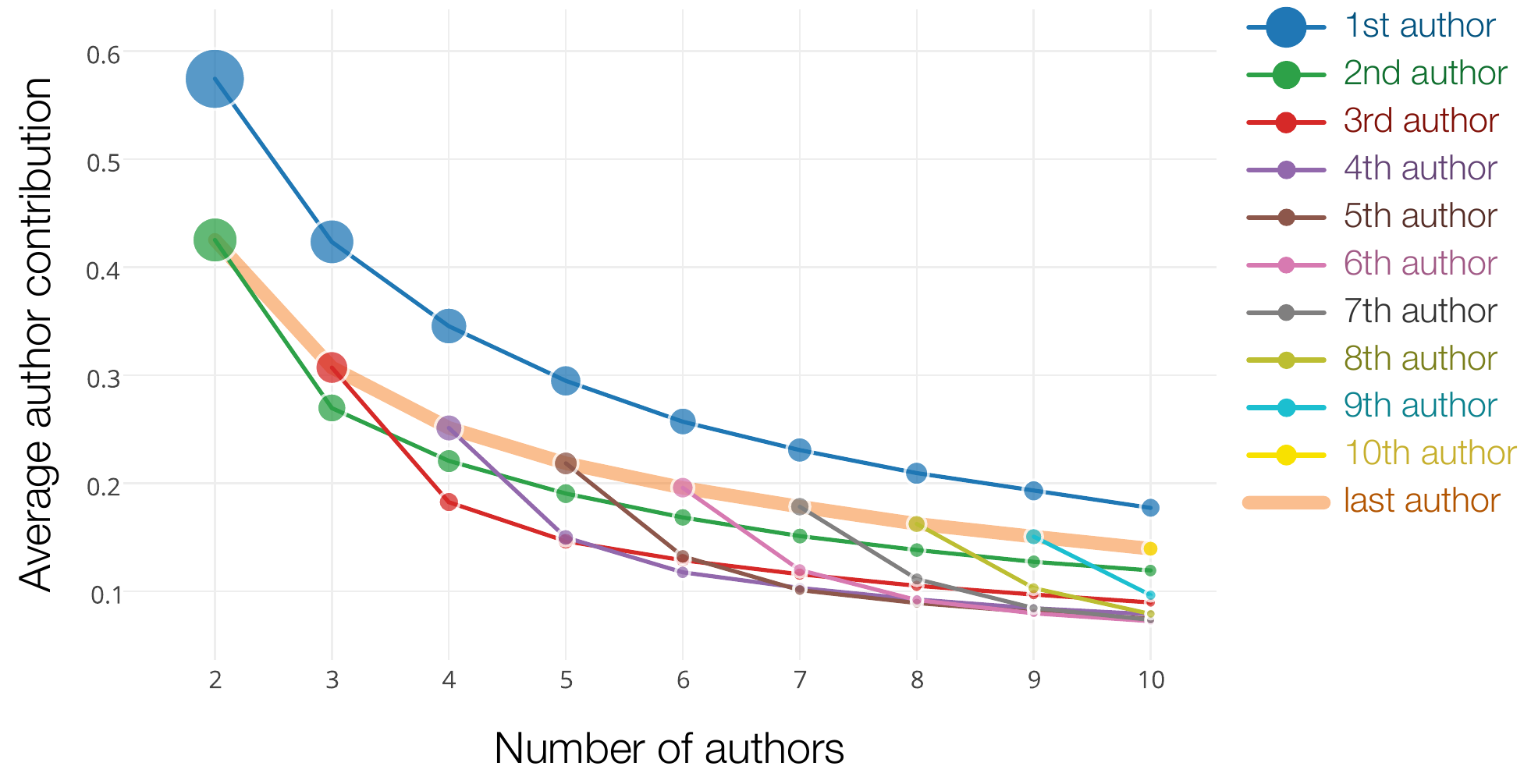}
\caption{\label{memedias}Average author contribution as a function of its ranking. In general, the first and last authors are the ones who most contribute to the paper (quantified in number of distinct contributions). }
\end{figure*}

In Figure \ref{contribution-multi}, we show the distribution of contributions made by each author according to their rank. The results are organized by the total number of authors considered in figure \ref{contribution-multi}, with papers authored by $2$ (a), $3$ (b), $4$ (c) and $8$ (d) authors (additional results are shown in the Supplementary Information, Figures S2 and S3). In Figure \ref{contribution-multi}(a), as expected~\cite{10.1371/journal.pone.0059814,10.1371/journal.pone.0084334}, it is evident that first authors, in general, make more contributions than last authors. Nonetheless, the amount of contributions are not very different, since, in average, first and last authors make, about 60\% and 40\% of the contributions, respectively. When more authors are included, one can observe a very similar pattern: while first authors make most of the contributions, last authors usually are ranked as the second most contributive authors.  Interestingly, disregarding first and last authors, the remaining ranking of authors reflects the amount of contributions made, i.e. second authors make more contributions than third authors, who in turn make more contributions than the fourth author and so forth. These patterns can also observed in Figure \ref{memedias}, which summarizes the average contribution made in terms of authors ranking. First authors (upper blue curve) always make most of the contributions, while last authors usually appear in the second position in the ranking of contributions. As the number of authors increases, however, there is not a large difference in contributions among the authors located from fourth to second to last positions.

\begin{figure*}[!htbp]
 \centering
   \includegraphics[width=0.95\linewidth]{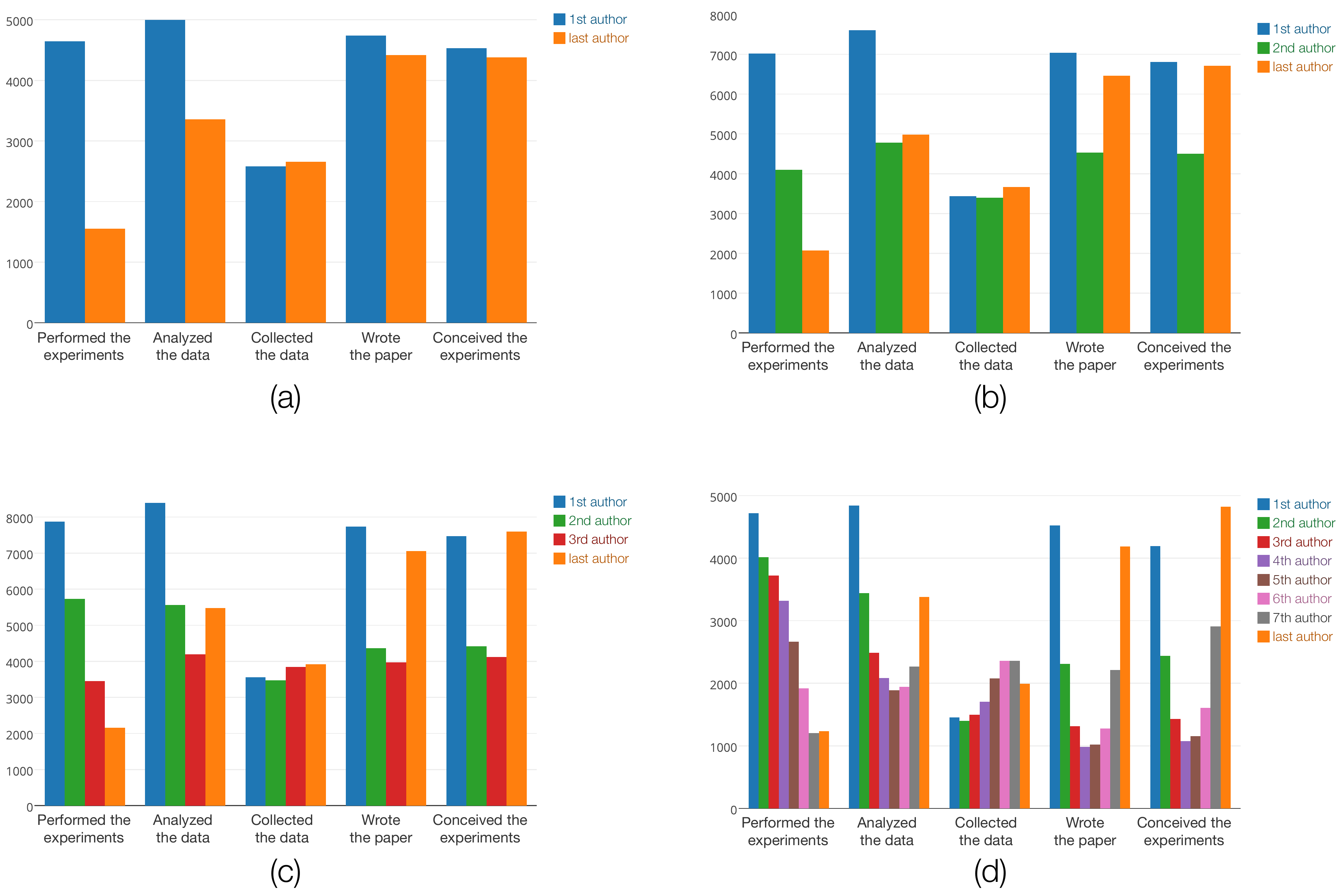}
   \caption{\label{contribution-multi-spec} Authors' contributions organized by type of contributions and ranks. The number of authors considered are: (a) 2 authors; (b) 3 authors; (c) 4 authors; and (d) 8 authors. Additional figures are shown in the Supplementary Information (Figures S2 and S3). The five most frequent contributions can be classified into three distinct patterns by taking into the account the relationship between amount of contribution made and authors rank.}
\end{figure*}

The relationship between contributions and rankings in authorship lists may not be clear when one analyzes intermediary rankings. It is conjectured that, in general, first authors are responsible for performing experiments, while last authors usually supervise the research. However, guidelines for ranking authors are not always strictly followed, and therefore there is no widespread evidence in relating ranking of intermediary co-authors  and specific contributions. To investigate the presence of patterns in ranking intermediary co-authors according to the type of their contributions, we show, in Figure \ref{contribution-multi-spec}, the total amount of authors in a particular ranking who made specific contributions. In Figure \ref{contribution-multi-spec}(a), we show that in papers authored by only 2 authors, both authors usually collect the data, write the paper and design the experiments in similar proportions. However, in most cases, first authors are responsible for performing the experiments, as one should expect. Interestingly, data analysis is mostly performed also by first authors.

Specific contributions made by authors in papers with 3 authors is shown in Figure \ref{contribution-multi-spec}(b). Note that, when comparing contributions of first and last authors, the proportions of contributions are very similar.
The intermediary author also makes an intermediary contribution in performing the experiments. Concerning the data analysis and acquisition, the contribution of intermediary and last authors are very similar. However, when considering paper writing and experiment design, the intermediary author usually make less contributions than main authors. Similar patterns of contributions have also been found for papers co-authored by 4 authors (see Figure \ref{contribution-multi-spec}(c)). In papers co-authored by many authors, three patterns of contributions could be identified (see e.g. Figure \ref{contribution-multi-spec}(d) or Figures S1 and S2 of the Supplementary Information):

\begin{enumerate}

\item {\bf Pattern A}: The total amount of author contributions \emph{decreases} with its ranking. This is the case of the contribution ``Performed the experiments'': first authors are the ones making most of this type of contribution, while last authors are usually the ones with the lowest contributions on preparing experiments.

\item {\bf Pattern B}:  The total amount of author contributions \emph{increases} from first to second-to-last authors. The last author performs an intermediary amount of contribution: he/she typically contributes more than first authors, but typically less than second-to-last authors. This pattern occurs for the contribution related to data collection.

\item {\bf Pattern C}: The total amount of author contributions displays a \emph{symmetric} behavior as a function of author ranking. From the first to the middle position, the amount diminishes; while from middle to last positions, the amount increases.  In other words, authors located in border positions (e.g. first, second, second-to-last and last positions) makes more contributions than intermediary authors. This type of behavior occurs for the following contributions: data analysis, manuscript writing and experiment design.

\end{enumerate}
The presence of three distinct patterns evidence that rankings and contributions are strongly related. These patterns also confirms that particular authors tend to make specific contributions. While most credit is devoted to first and last authors, here we show that intermediary authors may also contribute with a major frequency in contributions characterized by patterns A and B.

\section{Conclusion}\label{conclusion}

The authorship of articles has gained attention in recent years, mainly due to lack of protocols on how the authorship should be treated. It is well known that several journals make available guidelines on authorship, but even with this information the existence of gift authorships is still a problem~\cite{tscharntke2007author}. As the authorship in papers confers credit and has important academic, social and financial implications, a solution that has been adopted by several scientific journals is the identification of  the role of each author in preparing a scientific paper. Upon using this type of information, we studied several patterns of authors contribution in a large dataset recovered from the \emph{PLoS ONE} journal. We have first probed the symmetry of contributions to show that this quantity depends on the total number of authors. The symmetry/homogeneity of contributions obtained its highest average value in papers authored by 2 authors, and it diminished when the number of authors increased from 2 to 10. Interestingly, we have also found that there is no significant difference in the average symmetry of contributions when considering papers authored by more than 10 authors.

In this work, we have also found that there is a strong relationship between authors ranking and the total of contribution made by authors. We have confirmed that, in general, first and last authors make the most of the contributions to a scientific manuscript. Despite this well-known pattern, we have also identified that patterns of contributions can be described in a threefold manner, depending on the relationship between the amount of contributions and authors ranking. For example, intermediary authors are less probable to perform experiments when compared to first authors, however, they usually make more contributions in the experimental part than last authors. Another very interesting pattern concerns the collection of data for the study, as intermediary authors (especially those closer to the final positions in the authors list) usually contribute more than first authors.

As the information regarding individual contributions is provided in a growing number of journals, in future works, we intend to analyze how author profiles evolves along their career by analyzing patterns of contributions in scientific manuscripts. This same analysis could be perform in specific journals and areas, to better understand the relationship between contributions and ranking in specific research areas. This type of information could be useful to unveil novel patterns of credit assignment. However, perhaps the most important implication of studying  individual contributions in papers is its potential to improve the process of researchers evaluation through the creation of role-driven measures, which could be combined with traditional, well-established indexes such as the number of citations or the $h$-index.

\section*{Acknowledgements}
E. A. Corr\^ea Jr. and D. R. Amancio acknowledge financial support from Google (Google Research Awards in Latin America grant). D. R. Amancio also thanks S\~ao Paulo Research Foundation (FAPESP) for support (grant. no. 2014/20830-0). F. N. Silva acknowledges FAPESP (Grant No. 15/08003-4). L. da F. Costa thanks CNPq (Grant no. 307333/2013-2) and NAP-PRP-USP for support. This work has been supported also by FAPESP grant 11/50761-2.

\onecolumngrid

\newpage

\ \\ 

\newpage


%

\newcommand{\beginsupplement}{%
        \setcounter{table}{0}
        \renewcommand{\thetable}{S\arabic{table}}%
        \setcounter{figure}{0}
        \renewcommand{\thefigure}{S\arabic{figure}}%
     }
\clearpage

\beginsupplement

\appendix
\section{Supplementary Information}

\begin{figure*}[!htbp]
 \centering
   \includegraphics[width=0.85\linewidth]{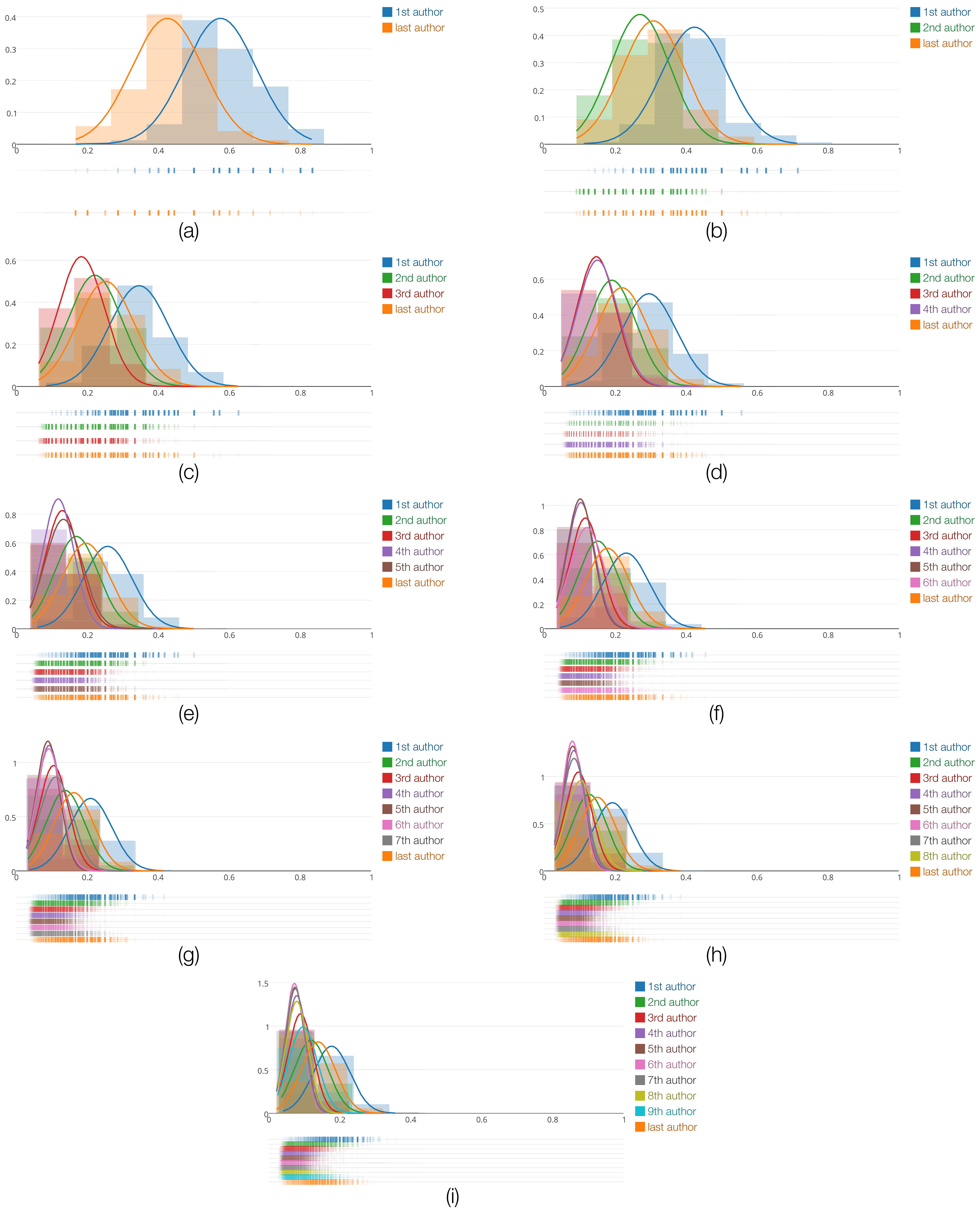}
\caption{Contributions of authors in scientific papers as a function of their rank. The results are shown considering the following number of authors: (a) 2; (b) 3; (c) 4; (d) 5; (e) 6; (f) 7; (g) 8; (h) 9 and (i) 10.}
\end{figure*}

\begin{figure*}[!htbp]
 \centering
   \includegraphics[width=0.95\linewidth]{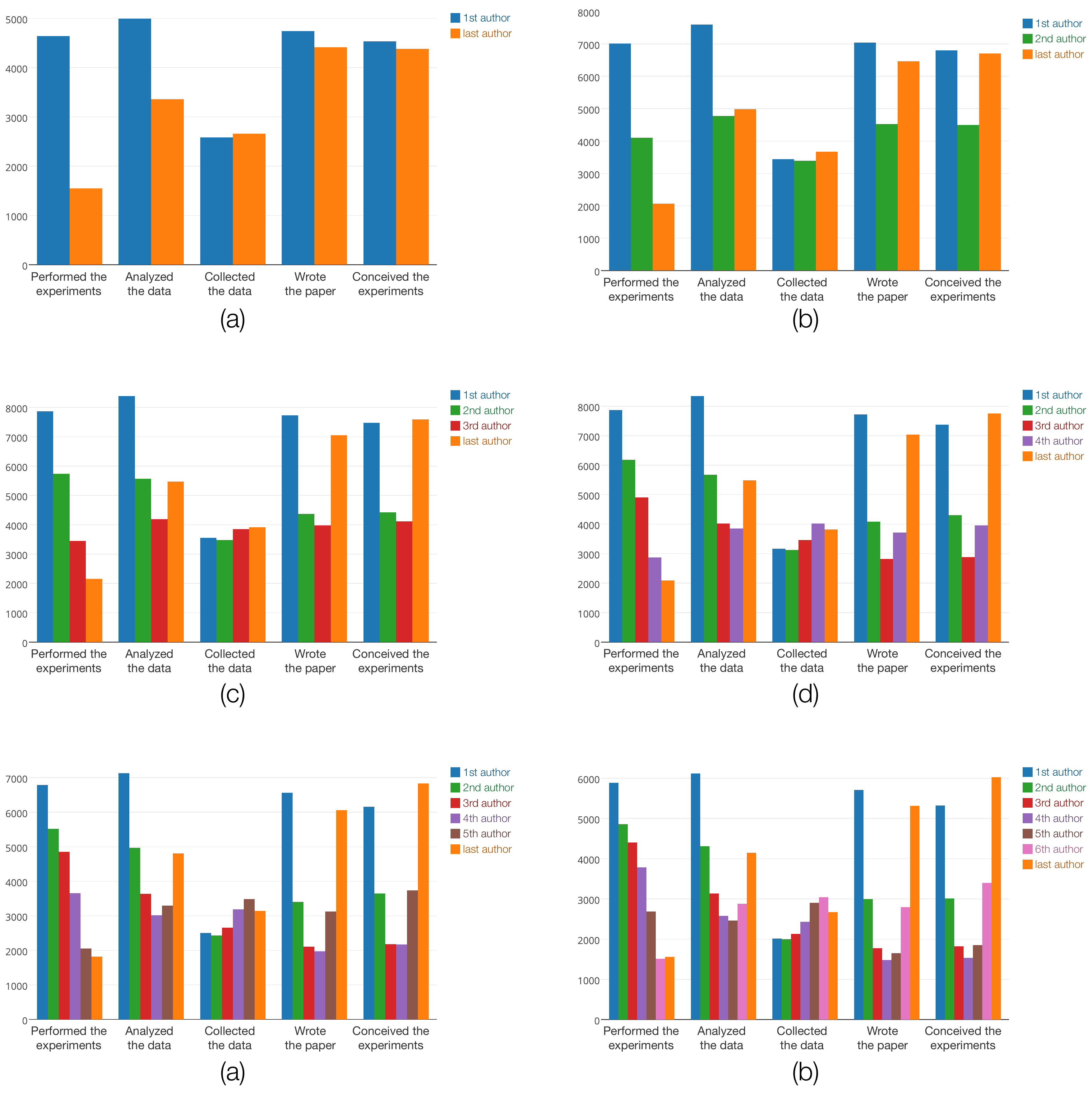}
   \caption{Authors' contributions organized by type of contributions and ranks. The number of authors considered are: (a) 2 authors; (b) 3 authors; (c) 4 authors; (d) 5 authors; (e) 6 authors; and (f) 7 authors. }
\end{figure*}

\begin{figure*}[!htbp]
 \centering
   \includegraphics[width=0.95\linewidth]{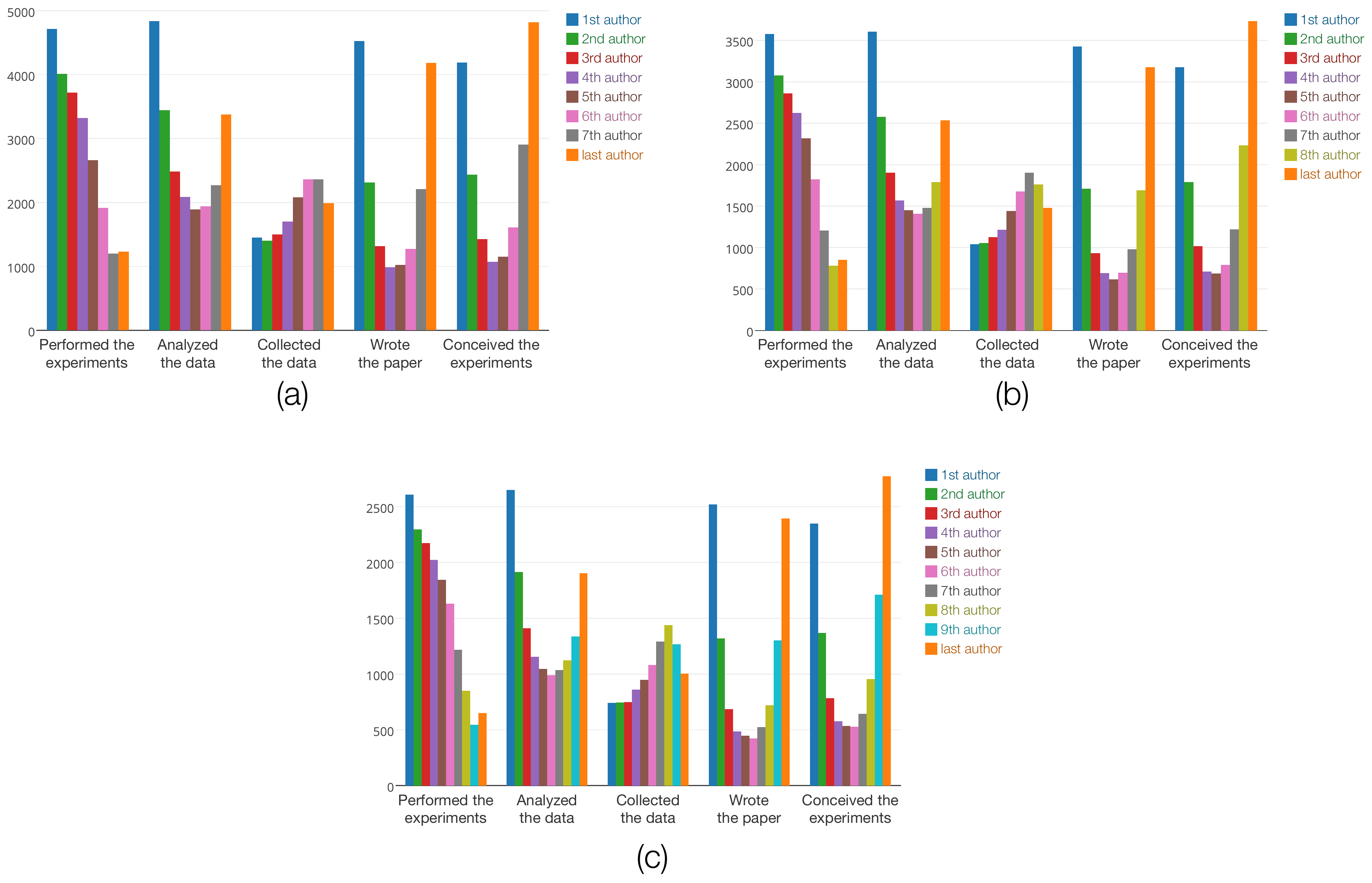}
   \caption{Continuation of Figure S2, displaying the authors' contributions for (a) 8 authors, (b) 9 authors and (c) 10 authors.}
\end{figure*}

\end{document}